# Sensitivity of propagation and energy deposition in femtosecond filamentation to the nonlinear refractive index

E. W. Rosenthal, J. P. Palastro*, N. Jhajj, S. Zahedpour, J.K. Wahlstrand, and H.M. Milchberg

*Institute for Research in Electronics and Applied Physics*

*University of Maryland, College Park, MD 20742*

*Icarus Research, Inc., P.O. Box 30780, Bethesda, Maryland 20824-0780*

**Abstract**

The axial dependence of femtosecond filamentation in air is measured under conditions of varying laser pulsewidth, energy, and focusing *f*-number. Filaments are characterized by the ultrafast *z*-dependent absorption of energy from the laser pulse and diagnosed by measuring the local single cycle acoustic wave generated. Results are compared to 2D+1 simulations of pulse propagation, whose results are highly sensitive to the instantaneous (electronic) part of the nonlinear response of $N_2$ and $O_2$. We find that recent measurements of the nonlinear refractive index ($n_2$) in [J.K. Wahlstrand *et al*., Phys. Rev. A. **85**, 043820 (2012)] provide the best match and an excellent fit between experiments and simulations.

1. Introduction

A high-intensity femtosecond optical pulse propagating through a gas deforms molecular electronic states and aligns molecules through impulsive excitation of rotational states [1, 2]. The resulting nonlinear polarization provides an intensity-dependent refractive index which causes the self-focusing spatial collapse of the pulse, with the intensity surpassing the ionization threshold. This generates free electrons concentrated on axis, whose optical response defocuses the pulse. The dynamic interplay between self-focusing and plasma-induced defocusing results in filamentary propagation [1], with a transversely localized and narrow (typically <100 μm diameter) on-axis region of high optical intensity, plasma, and atomic/molecular excitation whose axial extent greatly exceeds the Rayleigh range corresponding to its diameter.

Since the laser-induced atomic/molecular nonlinearity is responsible for the onset of filamentation and its sustainment, accurate coefficients are needed for modeling the nonlinear response in propagation models. Modeling and interpretation of experiments in filament-based applications such as long range propagation [3], high harmonic generation [4], and ultrashort pulse shaping and supercontinuum generation [5, 6], depend on an accurate representation of the nonlinearities. Many indirect measurements of the nonlinear response have appeared in the literature, with the aim of extracting coefficients such as $n_2$, the nonlinear index of refraction or Kerr coefficient [7]. Such indirect measurements include spectral analysis after nonlinear propagation [8], spatial profile analysis [9], polarization rotation by induced birefringence [10], and spectral shifts of a probe pulse [11, 12]. As an example, extracted $n_2$ values for the major constituents of air, $N_2$ and $O_2$, show a range of variation exceeding ~100%. Some of this variation can be attributed to nonlinear 3D propagation effects [8, 11], unintentional two-beam coupling in degenerate pump-probe experiments owing to the presence of laser-induced Kerr, plasma, and rotational gratings [10,11,13,14], and the laser pulsewidth dependence of the nonlinear response, which had not been directly resolved [8, 10-12].

At optical frequencies the electronic response, responsible for the Kerr effect, is nearly instantaneous on femtosecond time scales, while the response from molecular alignment is delayed owing to the molecular moment of inertia and depends strongly on the laser pulse duration [15,16,18]. The combined response can be expressed, to second order in the laser electric field, as a transient refractive index shift at a point in space,

$$\Delta n(t) = n_2 I(t) + \int_{-\infty}^{\infty} R(t-t')I(t')dt' \qquad , \qquad (1)$$

where $I(t)$ is the laser intensity, $R$ is the rotational Raman response function, and the first and second terms describe the instantaneous electronic and delayed rotational response. Experiments that use pulses longer than a few hundred femtoseconds [9,12] cannot distinguish the electronic from rotational response, making such results of limited use for understanding the propagation of ultrashort pulses. Even experiments using pulses that are 90-120 fs [8,10,11,18,19] are barely able to distinguish the two. Recently, the optical nonlinear response for a range of noble and molecular gases was absolutely measured using single-shot supercontinuum spectral interferometry (SSSI) using 40 fs pump pulses [15-17]. This measurement technique enabled accurate determination of the separate instantaneous and delayed contributions to the total response. A remarkable additional aspect of the measurements [15-17] is that the instantaneous part of the response is seen to be linear in the intensity envelope well beyond the perturbative regime all the way to the ionization limit of the atom or molecule. That is, the $n_2$ values measured in [16, 17] are valid over the full range of intensities experienced by atoms or molecules in the filament core.

Table 1 summarizes the results from these measurements for the major constituents of air, $N_2$, $O_2$, and Ar. Results from other experiments and calculations are shown for comparison, illustrating the wide range of values obtained.

|  | $n_2$ ($10^{-19}$ cm$^2$/W) | | | | | | $\Delta\alpha$ ($10^{-25}$ cm$^3$) |
|---|---|---|---|---|---|---|---|
|  | Wahlstrand et al. [16] (40 fs) | Nibbering et al. [8] (120 fs) | Loriot et al. [10] (90 fs) | Böerzsöenyi et al. [39] (200 fs) | Bukin et al. [40] (39 fs) | Shelton and Rice [41] | Wahlstrand et al. [16] |
| Air | 0.78 |  | 1.2 | 5.7 ± 2.5 | 3.01 |  |  |
| $N_2$ | 0.74 ± 0.09 | 2.3 ± 0.3 | 1.1 ± 0.2 | 6.7 ± 2.0 |  | 0.81 | 6.7 ± 0.3 |
| $O_2$ | 0.95 ± 0.12 | 5.1 ± 0.7 | 1.60 ± 0.35 |  |  | 0.87 | 10.2 ± 0.4 |
| Ar | 0.97 ± 0.12 | 1.4 ± 0.2 | 1.00 ± 0.09 | 19.4 ± 1.9 |  | 1.04 |  |

**Table 1.** Measured nonlinear coefficients for the major constituents of air. The Kerr coefficient, $n_2$, for the instantaneous atomic or molecular response, is shown from Wahlstrand *et al.* [16] with results from other experiments shown for comparison. Included are the pump pulse durations used in the measurements. Also shown is the molecular polarizability anisotropy $\Delta\alpha$, for which there is much less variability in the literature. The column for Shelton and Rice [41] gives semi-empirical results based on harmonic generation measurements at much lower laser intensity than in a filament core.

In this paper we explore the sensitivity of femtosecond filamentation in air to the nonlinear response of the constituent molecules. Experiments are performed with varying laser pulse energy, pulsewidth and focusing *f*-number, and filaments are diagnosed along their propagation path by evaluating the local energy density absorbed from the laser. The measurements are compared to laser propagation simulations in which the nonlinear coefficients pertaining to the instantaneous part of the response, namely the nonlinear indices of refraction $n_2$ for $N_2$ and $O_2$, are varied. We find sensitive dependence on the choices for $n_2$, with the best fit to experimental results obtained by using the values measured in [16]. For this sensitivity test, we focus on the instantaneous rather than the delayed response because of the prior wide variability in measured $n_2$, as displayed in Table 1. Our goal is to clearly demonstrate that accurate propagation simulations of high power femtosecond pulses depend sensitively on accurate values for the nonlinear response.

We have shown previously [20-22] that the ultrafast laser energy absorption during filamentation generates a pressure impulse leading to single cycle acoustic wave generation ~100 ns after the laser passes, followed at ~1 μs by a residual 'density hole' left in the gas after the acoustic wave propagates away. Hydrodynamics simulations show that for moderate perturbations to the gas, for which single filaments qualify, either the acoustic wave amplitude or the hole depth is proportional to the local laser energy absorbed [20-22]. While measurement of the density hole depth requires an interferometry setup with associated phase extraction analysis, the simplest approach is to measure the $z$-dependent acoustic amplitude with a microphone, and we use this signal as a proxy for laser energy absorbed by the gas.

Laser energy is nonlinearly absorbed by the gas through ionization and molecular rotational Raman excitation [22, 23]. (The bandwidth of typical ultrashort 800nm pulses is too small to support vibrational Raman absorption [18].) The rotational excitation thermalizes as the molecular rotational states collisionally dephase over a few hundred picoseconds [18], while the plasma recombines over ~10ns. Eventually, but still on a timescale much shorter than the fastest acoustic timescale of $a/c_s$ ~100 ns, where $a$ is the filament radius and $c_s$ is the sound speed, the absorbed laser energy is repartitioned over the thermodynamic degrees of freedom of the neutral gas and forms a pressure impulse that drives the subsequent hydrodynamics.

Acoustic measurements of optical filaments have been used in a number of prior contexts [24-27]. Other possible filament diagnostics are plasma conductivity [28], fluorescence [29], and direct [30] and indirect [29, 31-33] measurements of filament plasma density, none of which are directly proportional to absorbed laser energy, and all of which require a combination of non-trivial optical setups and data retrieval, and complex auxiliary modeling for interpretation.

## 2. Simulation of propagation and laser energy absorption

For the purposes of comparing the effects of different values of $n_2$ on filamentation, we employ a 2D+1 simulation of the optical pulse propagation [23, 34, 35]. The simulation models the most relevant aspects of the pulse's propagation, including the instantaneous electronic response, the delayed rotational response, multiphoton ionization, ionization damping, and the plasma response.

The transverse electric field envelope of the laser pulse evolves according to the modified paraxial wave equation

$$\left[\nabla_\perp^2 + 2\frac{\partial}{\partial z}\left(ik - \frac{\partial}{\partial \xi}\right) - \beta_2 \frac{\partial^2}{\partial \xi^2}\right] E = 4\pi \left(ik - \frac{\partial}{\partial \xi}\right)^2 P_{NL} \quad (2)$$

where $k = \omega_0 c^{-1}[1 + \delta\varepsilon(\omega_0)/2]$, $\omega_0$ is the pulse carrier frequency, $\delta\varepsilon(\omega)$ is the neutral gas contribution to the linear dielectric response, $\xi = v_g t - z$ is the position coordinate in the group velocity frame, $v_g = c[1 - \delta\varepsilon(\omega_0)/2]$, and $\beta_2/\omega_0 c = (\partial^2 k/\partial\omega^2)|_{\omega=\omega_0} = 20\,\text{fs}^2/\text{m}$ [36] accounts for group velocity dispersion in air. Included in the nonlinear polarization density, $P_{NL} = P_{elec} + P_{rot} + P_{free} + P_{ioniz}$, is the instantaneous electronic (Kerr) response, the delayed molecular rotational response, the (linear) free electron response, and a polarization density term associated with the laser energy loss from ionization (ionization damping).

It is convenient to express the electronic and rotational polarization densities as the product of an effective susceptibility and the electric field: $P_{elec} = \chi_{elec} E$ and $P_{rot} = \chi_{rot} E$, where

$$\chi_{elec} = \frac{1}{16\pi^2}\left(\frac{N_g}{N_{atm}}\right) n_2 |E|^2 \quad (3a)$$

$$\chi_{rot} = \sum_j \frac{N_g (\Delta\alpha)^2}{15\hbar} \frac{(j+1)(j+2)}{2j+3} \left(\frac{\rho^0_{j+2,j+2}}{2j+5} - \frac{\rho^0_{j,j}}{2j+1}\right) \int_{-\infty}^{\xi} \sin[\omega_{j+2,j}(\xi'-\xi)] |E|^2 d\xi' \quad . (3b)$$

Here, $n_2$ is the nonlinear index of refraction (Kerr coefficient) at 1 atm, $N_{atm}$ is the gas density at 1 atm, $N_g$ is the gas density, $\Delta\alpha$ is the difference in molecular polarizabilities parallel and perpendicular to the molecular bond axis, $j$ is the total angular momentum quantum number, $\omega_{j+2,j} = \hbar(2j+1)/I_M$, $I_M$ is the moment of inertia, and the $\rho^0_{j,j}$ are thermal equilibrium density matrix elements [18, 34].

The free electron polarization density is determined by $(ik - \partial_\xi)^2 P_{free} = (4\pi)^{-1} k_p^2 E$ where $k_p^2 = 4\pi e^2 N_e/m_e c^2$ is the square of the plasma wavenumber and $N_e$ is the free electron density. For inverse-Bremsstrahlung losses, we include an electron-neutral collision rate, $v_{en}$, on the left side of Eq. (2) when solving for the plasma response only. The ionization damping polarization density evolves as $(ik - \partial_\xi)P_{ioniz} = -2\kappa_{ion}E$, where

$$\kappa_{ion} = c^{-1}U_I \nu_I N_g \frac{1}{|E|^2} \quad (4)$$

is the damping rate, $U_I$ is the ionization potential, $\nu_I$ the cycle-averaged ionization rate [37], and $\partial_\xi N_e = c^{-1}\nu_I N_g$. A sum over species, namely nitrogen and oxygen, is implied in Eqs. (3) and (4). We neglect the contribution of Ar, which at ~1% atmospheric concentration has a negligible effect on the propagation simulation results.

With these expressions for the polarization densities and Eq. (2), the local depletion per unit length of the laser pulse energy, $U_L$, is given by

$$\frac{\partial}{\partial z}U_L = -\frac{1}{c}\int\left[2\pi\left(\frac{\partial \chi_{rot}}{\partial \xi}\right)I_L + \nu_I U_I N_g + c^{-1}\nu_{en}\left(\frac{\omega_p}{\omega_0}\right)^2 I_L + m_e c^2 K_{osc}\nu_I N_g\right]d^2rd\xi \quad (5)$$

where $I_L = (8\pi)^{-1}c|E|^2$ is the intensity and $K_{osc} = (e|E|/2m_e c\omega_0)^2$ is the normalized, cycle averaged quiver energy of a free electron. In order, the terms in the integrand represent the energy from the laser pulse absorbed (restored) by rotational excitation (de-excitation), the energy absorbed in freeing electrons from their binding potential (ionization energy), inverse-Bremsstrahlung losses, and the cycle-averaged kinetic energy imparted to electrons by the laser field as they enter the continuum, a result of freed electrons being born with zero velocity. This final term is often referred to as semi-classical above threshold ionization energy [38].

Our experiments use beam aperturing and weak focusing of the laser pulse to enable adjustment of the *f*-number and to promote filamentation. To model the aperture, a radial filter is applied to the electric field. In particular, the field just after the aperture, $E_{a,+}$, is given by $E_{a,+} = [1 - 4(r/r_a)^{18} + 3(r/r_a)^{24}]E_{a,-}$, where $r_a$ is the aperture radius and $E_{a,-}$ the field just before the aperture. We note that the filter function's value and derivative vanish at $r = r_a$. The lens is modeled by applying the thin-lens phase factor to the electric field $E_{l,+} = \exp[-ikr^2/2f]E_{l,-}$, where $E_{l,+}$ and $E_{l,-}$ are the fields just after and just before the lens and *f* is the lens focal length.

The laser input field is modeled as $E(\xi) = \sin(\pi\xi/\sigma)$ for $0 < \xi < \sigma$, where the FWHM of $|E(\xi)|^2$ is $\sigma/2$.

The simulations performed for this paper (see Sec. 4) examine the sensitivity of the axial profile of filament energy deposition to the choice of values of $n_2$ for N$_2$ and O$_2$. These determine the magnitude of the instantaneous part of the response and enter the simulation via Eq. (3a). The rotational response model, which is described by Eq. (3b) and uses the values of $\Delta\alpha$ from Table 1, remains unchanged for all simulations.

## 3. Experimental setup

The experimental setup is shown in Fig. 1. Pulses from a 10 Hz Ti:Sapphire laser system were apertured through a variable diameter iris immediately followed by a $f$ = 3m MgF$_2$ lens to gently initiate filamentary propagation. The pulsewidth, pulse energy, and iris diameter were varied while still producing stable single filaments. Single filament propagation was confirmed by visually inspecting the beam on an index card over the full range of propagation along the filament and examined in the far field. A compact electret-type microphone was mounted, 3mm away from the filament, on a rail to enable scans with ~1 cm axial resolution over the full maximum filament length of ~70 cm. The output signal of the microphone was digitized and collected by a computer for analysis. At each scan position along the filament, ~50 microphone traces were averaged. A typical average trace is shown in the figure. A CCD camera served as a shot-by-shot energy monitor using a small portion of the beam transmitted through a turning mirror. Energy binning allowed the discarding of laser shots deviating from the quoted pulse energy by more than ~10%. Note that the sound wave's maximum frequency is roughly $c_s/a$ ~ 10 MHz, greatly in excess of the microphone bandwidth's upper limit of ~15 kHz, so that the measured trace is simply the impulse response, whose peak is proportional to the acoustic wave amplitude.

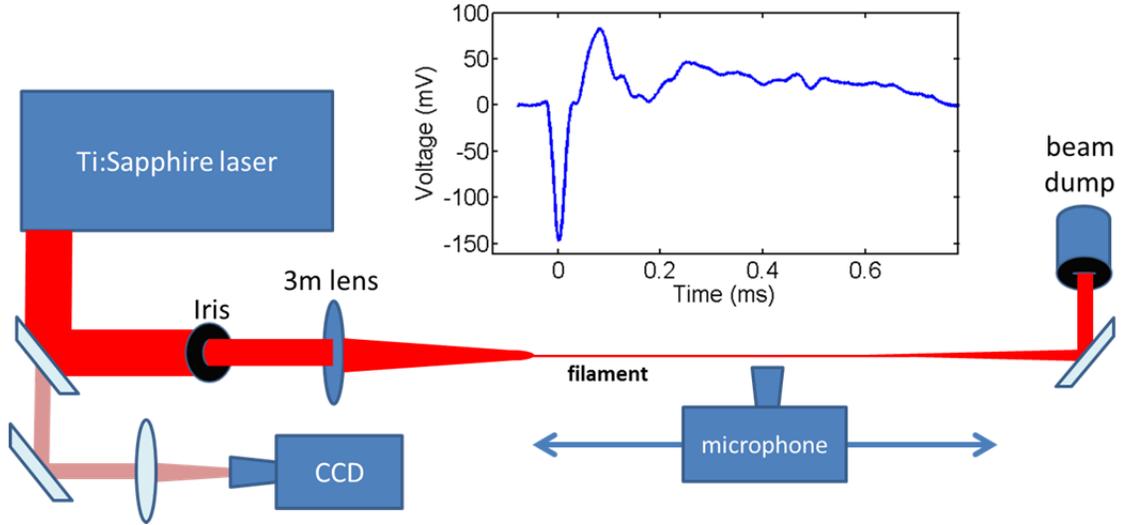

**Figure 1.** Experimental setup. Pulses from a 10Hz Ti:Sapphire laser are apertured by an iris and focused by a $f$=3m MgF$_2$ lens, forming an extended filament. A small portion of the laser energy is collected by a CCD camera to enable later energy binning of the results. An electret-type microphone positioned 3mm away from the propagation axis is axially scanned along the full length of the filament in 1 cm steps. Also shown is a typical averaged microphone signal.

## 4. Results and discussion

Figures 2 and 3 show microphone scans and propagation simulations for filaments generated with f/505 focusing (pulse energy 2.5 mJ) and f/300 focusing (pulse energy 1.8 mJ), for pulsewidths $\tau$=40 fs and $\tau$=132 fs. The pulse energy was reduced in the $f$/300 case to maintain single filamentation. As discussed above, the plotted points are proportional to the peak acoustic wave amplitude, which is proportional to the local energy absorption (or energy deposited per unit length) by the laser pulse. The simulation points are calculated as $-\partial U_L / \partial z$ from Eq. (5).

In the experiments, the laser pulsewidth was varied to explore the relative importance of choice of $n_2$ when filamentation is dominated by the instantaneous (Kerr) versus delayed (rotational) nonlinearities, and the $f$-number was varied to test the effect of lens focusing on the sensitivity of this choice.

Our prior work [15, 16, 30] has established that 40 fs pulses dominantly experience the instantaneous Kerr nonlinearity characterized by $n_2$, while the nonlinearity experienced by 132 fs pulses is dominated by molecular rotation. This is because the fastest onset timescale for the rotational contribution, $\Delta t_{rot} \sim 2T / j_{max}(j_{max}+1) > \sim 50$ fs, is set by the highest significantly populated rotational state $j_{max}$ (~16-18) impulsively excited in the filament at the laser pulse clamping intensity. Here $T$=8.3 ps is the fundamental rotational period for $N_2$ [18]. This leads us to expect that the choice of $n_2$ will be more significant for propagation simulations of shorter pulses.

We also expect that sensitivity to the choice of $n_2$ will be more pronounced in simulations of longer *f*-number-generated filaments. This is because larger *f*-numbers imply a weaker contribution of lens focusing, and a relatively more important role of nonlinear self-focusing to filament onset and propagation. For unaided filamentation of a collimated beam, the proper choice of $n_2$ in simulations is expected to be even more important.

Figure 2 shows experiment and simulation results for the longer *f*-number-generated filaments, at *f*/505. The left column of panels (green curves) is for 40 fs pulses and the right column of panels (red curves) is for 132 fs pulses. The experimental points are the same in each column, and the simulations explore the effect of using values of $n_2$ for $N_2$ and $O_2$ that are 50% (top row), 100% (middle row), and 150% (bottom row) of the measured values of Wahlstrand *et al*. [16] shown in Table 1.

In order to quantitatively assess the agreement between experiment and simulation, a two-dimensional $\chi^2$ fit test was performed according to

$$\left(\chi^2\right)_{jk} = N^{-1}\sum_{i=1}^{N}\left(\left(\lambda_j M(z_i) - S(z_i + \Delta z_k)\right)/(\lambda_j B(z_i))\right)^2 , \qquad (6)$$

where a scale factor $\lambda_j$ was applied to each set of *N* data points $M(z_i)$, and an axial shift $\Delta z_k$ was applied to each set of *N* points $S(z_i)$ simulating energy absorption. Here, $B(z_i)$ is the standard deviation of the mean corresponding to measurement $M(z_i)$. The scale factor $\lambda_j$ was adjusted over $10^4$ equally spaced values while $\Delta z_k$ was adjusted in 1 cm increments. The best fit is taken as $\chi^2 = \min\left((\chi^2)_{jk}\right)$, the minimum over *j* and *k*, and is shown on each panel of Fig. 2. In all cases, the optimum axial shift minimizing $\chi^2$ is less than 9 cm. It was separately verified that changing the effective focal length of the thin lens applied in the simulation by ~10cm does not change the shape of the simulated energy deposition; rather it changes the longitudinal position at which the energy deposition occurs.

It is seen in Fig. 2 that minimum $\chi^2$ is achieved for the middle row simulations using the values of $n_2$ for $N_2$ and $O_2$ given in Wahlstrand *et al*. [16]. For that case, the simulation curves match the experimental points surprisingly well.

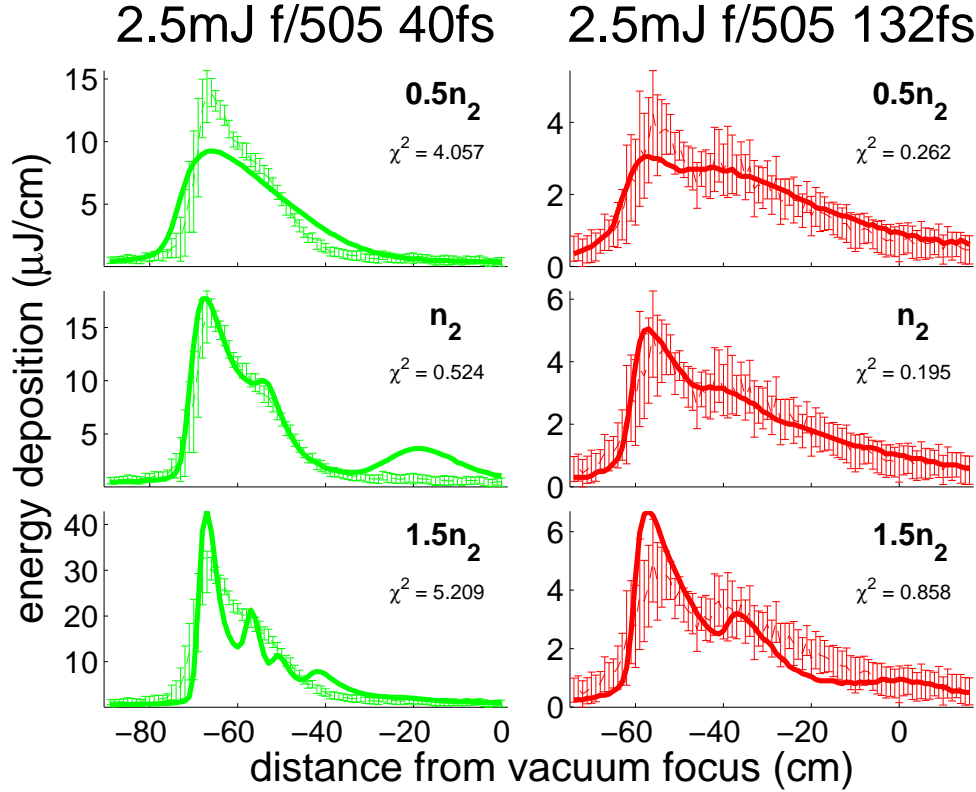

**Figure 2.** Axial scan of average peak signal from microphone trace (points) and propagation simulations of laser energy deposition (solid curves). Filaments were generated with pulse energy 2.5 mJ at *f*/505 for pulsewidths 40fs (green) and 132fs (red). The error bars on the points are the standard deviation of the mean for ~50 shots at each axial location. The simulations in the center row use $n_2$ values for $N_2$ and $O_2$ from Wahlstrand *et al*. [16] (see Table 1), while simulations in the top and bottom rows use 1.5 times and 0.5 times these values. The vacuum focus position is z=0. The $\chi^2$ fit result is shown on each plot.

Further examination of Fig. 2 shows that the experiment-simulation mismatch in the shorter pulse (40 fs) case is more sensitive to the choice of $n_2$ than in the longer pulse case (132 fs). As discussed earlier, the reason for this is that the dominant positive nonlinearity governing propagation in the long pulse case is field-induced molecular rotation, with reduced sensitivity to the instantaneous response characterized by $n_2$. It is worth noting that in the long pulse case, the signal does not go to zero at either end of the plot because measurable filament energy deposition extended beyond the range of the microphone rail travel.

Results from experiments and simulations for filaments generated at a lower $f$-number, $f/300$, are shown in Fig. 3. The figure panels are organized in the same way as in Fig. 2. Here again, it is seen that the best fit between simulation and experiment, as measured by $\chi^2$, is for simulations using the $n_2$ values measured in Wahlstrand *et al*. [16]. These simulations match the experiment quite well. There are two additional important observations. First, as before, and for the same reason, the long pulse (132 fs) simulations are less sensitive to choice of $n_2$ than short pulse simulations. Second, even with the greater sensitivity of the short pulse simulations to choice of $n_2$, that sensitivity is reduced from the $f/505$ case of Fig. 2. This is because at $f/300$ (which induces ~70% more phase front curvature), the lens plays a relatively more important role in the filament propagation.

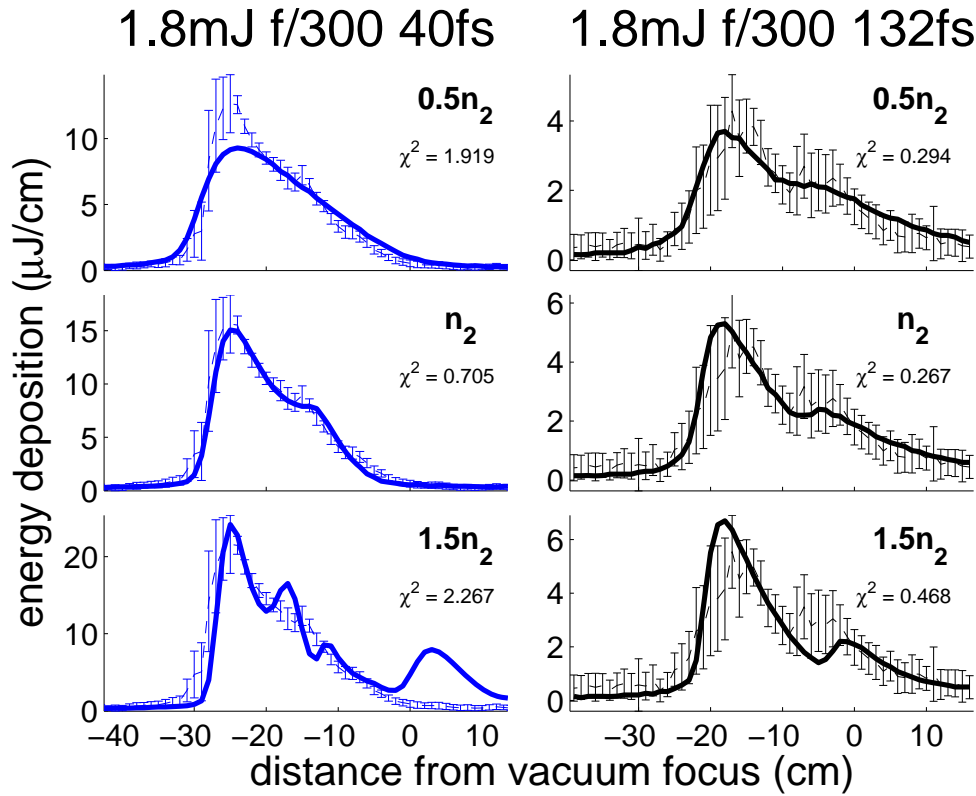

**Figure 3.** Axial scan of average peak signal from microphone trace (points) and propagation simulations of laser energy deposition (solid curves). Filaments were generated with pulse energy 1.8 mJ at $f/300$ for pulsewidths 40fs (blue) and 132fs (black). The error bars on the points are the standard deviation of the mean for ~50 shots at each axial location. The simulations in the center row use $n_2$ values for $N_2$ and $O_2$ from Wahlstrand *et al.* [16] (see Table 1), while simulations in the top and bottom rows use 1.5 times and 0.5 times these values. The vacuum focus position is $z=0$. The $\chi^2$ fit result is shown on each plot.

There are several locations in the short pulse simulations (middle green panel of Fig. 2 and bottom blue panel of Fig. 3) showing a downstream resurgence in the laser absorption. This is an artifact produced by the radial symmetry assumed by the simulation, which arises due to a combination of space-time focusing and plasma refraction at the back of the pulse. Azimuthal intensity variation in real experimental beam profiles (and the associated azimuthally varying nonlinear phase pickup) significantly reduces the on-axis superposition of beam contributions, thereby reducing or eliminating the energy deposition compared to the simulation.

Our simulations also allow an examination of the individual laser absorption channels in air. Figure 4 shows plots of three of the absorption terms in Eq. (5). The contribution of inverse bremsstrahlung absorption (third term) is negligible because $\nu_{en}\tau \ll 1$, and is not shown. Two contributions dominate for most of our measured filament parameters. One is rotational excitation of molecules, described by the first term in Eq. (5). The other dominant channel is the energy absorbed in ionization, here taken as the promotion of bound electrons to the continuum with zero velocity (second term in Eq. (5)). The contribution of the fourth term, the excess energy from above threshold ionization (modeled as a semi-classical electron drift energy), which goes into electron heating, is comparatively less significant. For lower intensity and longer duration pulses, the molecular rotation channel can dominate ionization, as seen for a large portion of the filament length in the *f*/505, $\tau$=132 fs case. This follows straightforwardly from the reduced ionization rate at lower intensity and the more efficient coupling to molecular rotation of longer pulses [16, 18, 30]. Conversely, for higher intensity pulses, ionization dominates. It is interesting to note that in both cases of Fig. 4, there is significant molecular absorption both in advance of the onset of ionization and well beyond where ionization fades away.

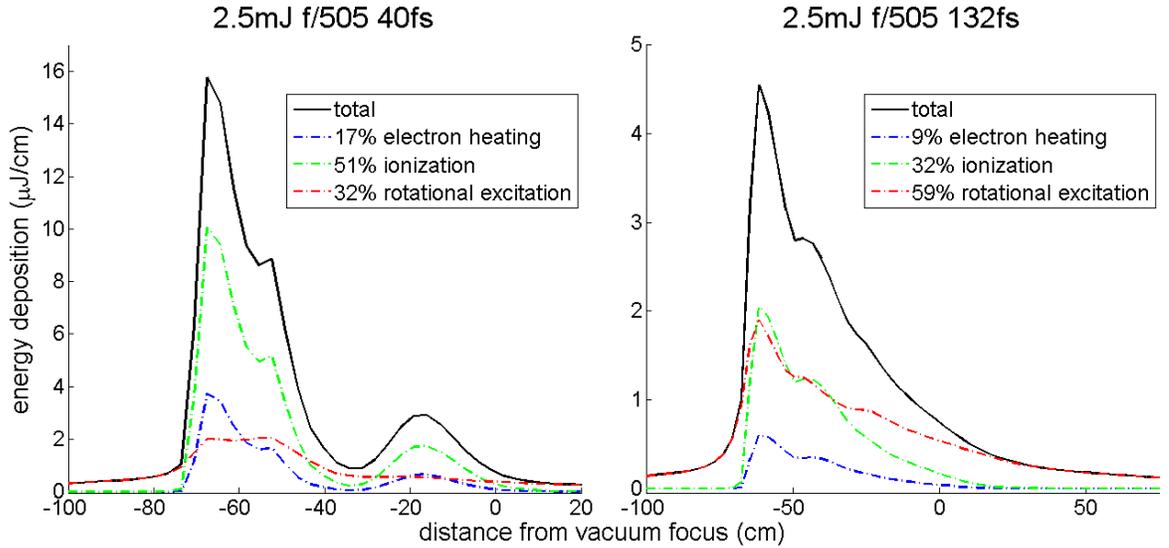

**Figure 4.** Simulated energy deposition due to various mechanisms in air for the laser parameters shown above each panel. The solid curve (black) represents the total energy deposited into the air, while dotted curves represent the energy deposited through above threshold ionization (blue), ionization of the medium (green), and rotational excitation (red). Inverse bremsstrahlung heating of the electrons is negligible and not shown.

## 5. Conclusions

We have shown that the *z*-dependent monitoring of the acoustic wave launched by a filament is a remarkably sensitive diagnostic of the laser energy absorption physics of filamentation. This diagnostic has enabled detailed comparisons of filament propagation experiments with simulations. It was seen that simulations of filament propagation in air depend sensitively on the choice of the nonlinear indices of refraction, $n_2$, which describe the instantaneous portion of the nonlinear response. The values of $n_2$ for $N_2$ and $O_2$ providing the best fit between simulation and experiment are those measured in Wahlstrand *et al*. [16], with excellent agreement in that case. For longer laser pulses and lower *f*-number induced filamentation, sensitivity to the proper choice of $n_2$ is somewhat reduced due to the relatively larger roles of the molecular rotational nonlinearity and the lens focusing. Based on our results, we expect that the most sensitive test for the proper choices of $n_2$ is beam collapse and filamentation by a collimated beam without assistance from a lens.

**Acknowledgements**

This work is supported by the Air Force Office of Scientific Research, the Army Research Office, the Defense Threat Reduction Agency, and the National Science Foundation.